\documentclass[nofootinbib,twocolumn,preprintnumbers]{revtex4-1}
\usepackage[dvips]{graphicx}
\usepackage{amsmath,amsthm,amssymb,multirow,psfrag}

\def\gev{\; \hbox{GeV}}

\def\tta{\theta^*}
\def\ifb{\; \hbox{fb}^{-1}}
\def\to{\rightarrow}

\begin{document}

\title{Is the New Resonance Spin 0 or 2? Taking a Step Forward in the Higgs Boson Discovery
\bigskip{}
}



\author{Alexandre Alves}

\affiliation{Universidade Federal de S\~ao Paulo, UNIFESP,  Dep. de Ci\^encias Exatas e da Terra, Diadema-SP 09972-270, Brazil}

\begin{abstract}
The observation of a new boson of mass $\sim 125\gev$ at the CERN LHC may finally have revealed the existence of a Higgs boson. Now we have the opportunity to scrutinize its properties, determining its quantum numbers and couplings to the standard model particles, in order to confirm or not its discovery. We show that by the end of the 8 TeV run, combining the entire data sets of ATLAS and CMS, it will be possible to discriminate between the following discovery alternatives: a scalar $J^P=0^+$ or a tensor $J^P=2^+$ particle with minimal couplings to photons, at a $5\sigma$ statistical confidence level at least, using only diphotons events. Our results are based on the calculation of a center-edge asymmetry measure of the reconstructed {\it sPlot} scattering polar angle of the diphotons. The results based on asymmetries are shown to be rather robust against systematic uncertainties with comparable discrimination power to a log likelihood ratio statistic.
\end{abstract}
\maketitle

\section{Introduction}

The CERN LHC has recently achieved one of its main goals -- the discovery of a new particle whose known properties, until this moment, points to a Higgs boson~\cite{atlas,cms}. In fact, as the new particle decays to pairs of gauge bosons and fermions, a non-integer spin is ruled out. Moreover, the new resonance decays into pairs of photons which, accordingly to the Landau-Yang theorem~\cite{landau-yang}, discards the spin-1 possibility, leaving either a scalar or a tensor option for the new boson.

Concerning the parity properties of the new boson, as a pseudo-scalar has no renormalizable interactions to SM particles, and given the lack of evidence of new physics at the electroweak scale, the $J^P=0^-$ possibility is currently disfavored by collider data, noticeably the $H\to WW$ branching ratio which is expected to be two orders of magnitude smaller than the observed one~\cite{logan}. 

The measured mass, production cross section, and branching ratios favor the Higgs boson hypothesis though. In fact, scalar impostors alternatives can also explain the observed signal~\cite{impostors} while spin-2 alternatives like KK-gravitons from ADD~\cite{add} or Randall-Sundrum (RS)~\cite{rs} models predict different decay patterns at the same time they are already very constrained by the LHC data~\cite{gravsearch}. Nevertheless, a dedicated study of the spin of the new resonance is a necessary step to ascertain its scalar or tensorial nature and, consequently, fulfill a necessary condition to claim the discovery of a Higgs boson.


The possible enhancement in the diphoton channel has motivated several works of Standard Model (SM) extensions capable to explain the experimental observation~\cite{diphotons} although a SM Higgs boson is by no means ruled out~\cite{tilman,model_indep} by the present data. Regardless the absolute size of the branching ratio to photons, all the proposed SM extensions that involve a scalar particle predict an isotropically distributed yield at parton level. This is not the case for a typical spin-2 state which decays in a highly anisotropic way inside the detector~\cite{spin_gravitons}. 

Analyzing the shape of the angular variables of charged leptons, with multivariate likelihood techniques, proved to be able to discriminate between a scalar, a vector, or a tensor hypothesis for a favorable mass region in $ZZ\to \ell_1^+\ell_1^-\ell_2^+\ell_2^-$ decays with sufficient data at an upgraded 10 or 14 TeV LHC~\cite{zerwas,lykken,gritsan1}. 

More recently, and update of the $ZZ$ channel, and a brand new analysis for spins in the $WW$ and $\gamma\gamma$ channels were made for the 8 TeV LHC~\cite{gritsan2}. Concerning the spin-0 {\it versus} spin-2 discrimination, the main conclusion of this work was that around a $4\sigma$ confidence level in the CP-even spin-0 resonance hypothesis is expected for $35\ifb$ at one LHC experiment, combining the $ZZ$, $WW$, and $\gamma\gamma$ channels.

Spin analyses for spin-2 RS KK-gravitons at the LHC, relying on asymmetries of angular distributions, were carried out in dilepton and diphoton channels based on $\chi^2$-tests~\cite{spin_gravitons}. As the branching ratio of these KK-gravitons into photons is much larger than the corresponding SM Higgs boson, the backgrounds are not a serious issue for spin discrimination in this case.

 The experimental fact, however, is that the $\gamma\gamma$ channel alone does not look promising given the tiny observed signal to background ratio of ${\cal O}(10^{-3})$ in the mass region of $100\gev$--$160\gev$~\cite{atlas,cms}. Whatever the kinematic variable chosen to discriminate between the two hypothesis, the huge background contamination would wash away any distinctive feature of the distributions in this channel. Requiring harder acceptance cuts, on the other hand, may help to clean the data sample, but at the cost of modifying the observed angular distributions which are used to compare the predictions of different spin hypotheses. 

Therefore, determining the spin of the new resonance studying the shape of the kinematic distributions of photons would require a truly clamping technique to pick the signal events from the huge ensemble of background events without demanding too hard cuts. This is precisely what has been suggested for the $ZZ$ channel~\cite{lykken} by means of the {\it sWeight} technique~\cite{splot}. In this channel, signal to background ratios are also very small, and an optimal separation of signal and background angular distributions was achieved by using {\it sPlots}~\cite{splot}. The technique has also been used in several experimental studies at BaBar, Tevatron, LHCb, and the LHC~\cite{exps}. 

 We show that, weighting diphoton events according to their well understood invariant mass distributions using the {\it sWeight} technique, a high confidence level discrimination is possible between a scalar $J^P=0^+$ and a tensor $J^P=2^+$ resonance with minimal couplings to SM particles. This is done by measuring a center-edge asymmetry~\cite{ace} of the polar angle of the photons in the resonance rest frame. At the time when the $\gamma\gamma$ excess reaches a $5\sigma$ significance level in a SM-like Higgs boson signal against backgrounds, a spin-0 or spin-2 will be favored to an even higher confidence level. We also show that the asymmetry is rather insensitive to systematic errors and can give even better results than a log likelihood ratio analysis. 

The paper is organized as follows: in section II we present the expected parton and detector level distribution for the scattering polar angle of diphotons  in spin-0 and spin-2 decays. In section III we summarize the {\it sWeight} technique. The section IV is devoted to computation of the statistical significances based on the asymmetry measure and the log likelihood ratio statistic including systematic uncertainties. Results are presented in section V and the section VI contains our conclusions.

\section{Polar Angle Distribution of Diphotons from Spin-0 and Spin-2 Resonance Decays}

 In this section we present the parton and hadron/detector level distributions for the scattering production angle $\tta$ in the center-of-mass (CM) system of the photon pair at the LHC. 

\subsection{Parton level distributions}

We concentrate on the main production mode of a Higgs boson at the LHC, the gluon fusion process~\cite{djouadi}, with subsequent decay to a pair of photons
\begin{equation}
g(\lambda_1^\prime)g(\lambda_2^\prime)\to X_0,X_2\to \gamma(\lambda_1)\gamma(\lambda_2)
\label{ggtoaa}
\end{equation}
Here $\lambda_{1,2}=\pm 1$ denote the photon helicities, $\lambda_{1,2}^\prime=\pm 1$ the gluon helicities, and $X_0,X_2$ a spin-$0^+$ and spin-$2^+$ resonance, respectively.

 In the resonance rest frame, the theoretical parton level $\tta$-distribution for a spin-0 resonance decaying to two photons is independent of the $\tta$ angle
\begin{equation}
\frac{d\Gamma_0}{d\cos(\tta)}\propto P_0(\tta)=|d^1_{0,0}(\tta)|^2=1
\label{P0}
\end{equation}
where $P(\tta)$ denotes the probability density function (pdf) of the $\tta$ variable, and $d^J_{m,m^\prime}(\tta)$ are the Wigner functions~\cite{pdg}.

The amplitude for the decay of a spin-2 state of mass $m_G$  into two photons can be conveniently written as~\cite{gritsan1,gritsan2}
\begin{equation}
{\cal M}(X_2\to \gamma\gamma) = \frac{g_1}{\Lambda}{\cal O}_1+\sum_{i=2}^5\frac{g_i}{\Lambda^3}{\cal O}_i
\label{coup}
\end{equation}
where ${\cal O}_1$ is a dimension-2 operator, while ${\cal O}_i$, $i=2,3,4,5$, are dimension-4 operators, and $\Lambda$ an energy scale where new physics is expected to appear. The minimal coupling of a spin-2 state to photons corresponds to set $g_1\neq 0$ and $g_{i\neq 0}=0$. The explicit expressions of these operators can be found in the appendix and in Refs.~\cite{gritsan1,gritsan2} as well.

The helicity amplitudes of the decay $X_2\to \gamma(\lambda_1)\gamma(\lambda_2)$ can be expressed in terms of the $g_i$ couplings of Equation~(\ref{coup}) as
\begin{eqnarray}
{\cal M}(++) &=& \frac{m^2_G}{\sqrt{6}\Lambda}\left(\frac{c_1}{2}+2c_2+i(c_3-2c_4)\right) \nonumber \\
{\cal M}(--) &=& \frac{m^2_G}{\sqrt{6}\Lambda}\left(\frac{c_1}{2}+2c_2-i(c_3-2c_4)\right) \nonumber \\
{\cal M}(+-) &=& \frac{m^2_G}{2\Lambda}c_1 \nonumber \\
{\cal M}(-+) &=&  {\cal M}(+-)
\label{helicity}
\end{eqnarray}
Here, the $c_i$ are combinations of $g_i$ couplings: $c_1=2g_1+g_2\kappa$, $c_2=-g_1/2+\kappa (g_3/2+g_4)$, $c_3=(-g_2/2+g_3+2g_4)\kappa$, $c_5=2g_5\kappa$, and $\kappa=m^2_G/\Lambda^2$~\cite{gritsan1,gritsan2}.

 The amplitudes of Equation(\ref{helicity}) take into account the most general couplings between a spin-2 state and massless vectors. In this point, we specialize our analysis to the minimal couplings between the spin-2 resonance and photons and gluons. It has been suggested that some non-minimal couplings configurations are already disfavored by the current data~\cite{petriello}. Note that setting $g_5\neq 0$ and $g_{i\neq 5}=0$ corresponds to the pure parity-odd tensor state $2^-$.

If $g_4$ or $g_5$ are not vanishing, for instance, the spin-0 {\it vs} spin-2 discrimination is an easier task in the $\gamma\gamma$ and $ZZ$ channels if we explore the angular distributions of photons and charged leptons, respectively~\cite{gritsan2}. In this respect, discriminate between a $0^+$ and a $2^+$ hypothesis is not the easiest conceivable scenario. Yet, through a careful tuning of the $g_i$ couplings, is possible to approximately mimic the $\cos(\tta)$ distribution of a spin-$0^+$ state decaying to photons.

 Our main goal, in this work, is to show that the {\it sWeight} technique, for which we give a brief description in the next section, is a powerful tool to study the spin of the new resonance in the $\gamma\gamma$ channel. We choose the minimal couplings scenario for a resonant spin-2 particle as a concrete example. 


A common feature of theoretically well motivated extra-dimensions models, as the large extra dimensions scenario of the ADD model~\cite{add} and the warped extra dimension Randall-Sundrum model~\cite{rs}, is the existence of spin-2 Kaluza-Klein gravitons. The collider effects of large extra dimensions of the ADD scenario are expected to be observable as towers of almost degenerated KK-gravitons~\cite{rizzo}. Despite it might be possible to study the angular dependence of photons coupled to such KK-towers, the {\it sWeight} analysis requires a well understood $\gamma\gamma$ invariant mass distribution to work. This is more easily accomplished for narrow resonances\footnote{Narrow resonances of Regge excitations with minimal couplings to SM particles are also expected in open string theories~\cite{anchor}.} as those expected from the RS warped extra-dimension scenario with large mass splittings among the graviton states~\cite{rizzo}.

 The Lagrangian for the interaction between SM particles and a massive RS KK-graviton in 4 space-time dimensions is given by
\begin{equation}
{\cal L}_{int}=-\frac{1}{\Lambda}G_1^{\mu\nu}T_{\mu\nu}
\label{kklag}
\end{equation}
where $G_1^{\mu\nu}$ is the field of the $1^{st}$ massive KK-graviton mode, $T_{\mu\nu}$ the SM stress-energy tensor, and $\Lambda$ the scale up to the effective theory is valid and that can be identified with the energy scale in Equation~(\ref{coup}). This interaction Lagrangian involves only  minimal couplings between the graviton and SM fields~\cite{taohan,strings, ellis}.

With minimal couplings, $c_2=-4c_1$ and all the other $c_i$ vanish. As the result of an accidental cancellation, ${\cal M}(++)={\cal M}(--)=0$, and the decay width of the KK-graviton to photons results in the following formula
\begin{eqnarray}
\frac{d\Gamma_2}{d\cos(\tta)} & \propto &  P(\tta)\propto \sum_{\lambda_1,\lambda_2}
\sum_{m} \lvert\langle 11,\lambda_1\lambda_2\lvert S\rvert 2,m\rangle\rvert ^2\nonumber \\
&=& \sum_{\lambda_1,\lambda_2}\sum_m \frac{5}{4\pi}\lvert {\cal M}(\lambda_1\lambda_2)\rvert^2\lvert d^2_{m,\lambda_1-\lambda_2}(\tta)\rvert^2 \nonumber \\
&=& \frac{5}{4\pi}\lvert {\cal M}(+-)\rvert^2\left(\lvert d^2_{2,2}(\tta)\rvert^2+\lvert d^2_{2,-2}(\tta)\rvert^2\right)\nonumber \\
& &
\label{width}
\end{eqnarray}

 The Equation~(\ref{width}) reproduces the well known result~\cite{ellis,gritsan2,oscar} for the normalized $P_2(\tta)$
\begin{equation}
P_2(\tta) = \frac{5}{32}\left(1+6\cos^2(\tta)+\cos^4(\tta)\right)
\label{P2grav}
\end{equation} 

Despite it is not favored by the current data in the Higgs search and by the stringent constraints from resonant production and decay to $WW$ bosons~\cite{gravsearch}, we take the Randall-Sundrum model, where a narrow spin-2 resonance couples minimally to the SM particles, as a straw man scenario against with the spin-0 hypothesis can be tested. This is a mere convenience. We could also have used a similar model with identical minimal couplings with no reference to any known model. In particular, we normalize all the spin-2 cross sections to the SM Higgs boson ones, leaving the $\cos(\tta)$ distribution as the only discriminant. In the following discussions we will refer also to the spin-2 resonance as a KK-graviton.

We show in the Figure~(\ref{dcos}) the parton level distributions of the $\cos(\tta)$ variable for a Higgs boson and a KK-graviton with minimal couplings according to Equations~(\ref{P0}) and~(\ref{P2grav}). These distributions are considerably distorted by detector acceptances as we discuss in the next section.

\subsection{Distorted shapes including detector effects}
\begin{figure}
\resizebox{0.5\textwidth}{!}{%
\includegraphics[height=.20\textheight]{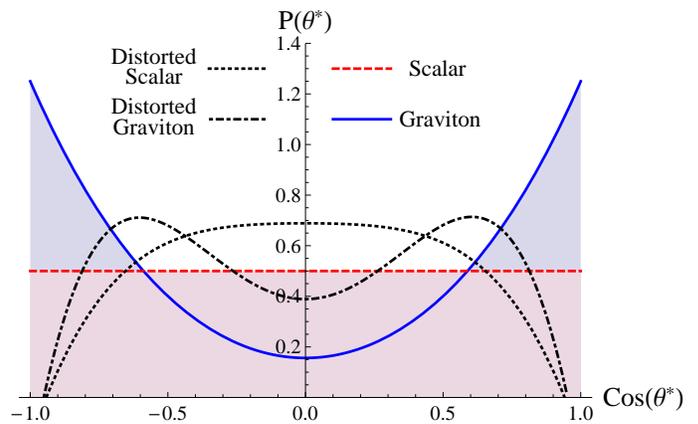}}
\caption{The normalized pdf of the angular variable $\cos(\tta)$. The solid blue line represents the parton level predictions for a spin-2 KK-graviton, and the flat dashed line a spin-0 distribution. The dotted line shows the distorted distribution after including detector effects for hadron level events for the spin-0, and the dot-dashed line the distorted spin-2 KK-graviton case.}
\label{dcos}
\end{figure}
The parton level distributions are considerably distorted after taking into account initial-state radiation  and detector resolution effects. The geometric restrictions of the detectors, reflected in the finite rapidity coverage, play the role of boundary conditions on the $P(\tta)$ functions such that $P(\tta=\tta_c,\pi-\tta_c)=0$ where $\tta_c$ is the limiting angle from the rapidity cut. 

In order to simulate these effects in the shape of the $\tta$ distributions, we generated a large number of events (typically ${\cal O}(10^5)$) to a SM Higgs boson, to represent the $0^+$ case; a massive first excited state of a RS KK-graviton, representing the $2^+$ alternative; and the irreducible and main reducible backgrounds for $\gamma\gamma$ production at the LHC, using \texttt {MadGraph5}~\cite{mad5} with CTEQ6L1 parton distribution functions~\cite{cteq}. The parton level events were, then, interfaced to \texttt{Pythia 6.4}~\cite{pythia} to simulate hadronization and jet clustering, and to \texttt{PGS} to a fast detector simulation with ATLAS settings~\cite{atlas}.

The hard matrix element contributions and soft radiation from parton showers were consistently merged in the MLM scheme~\cite{mlm}. The merging is crucial for a reliable graviton simulation because there are tree level diagrams contributing to the inclusive process $pp\to\gamma\gamma + X$ through $q\bar{q}gG$, $gggG$ and $ggggG$ vertices. Despite these are not leading contributions, their impact on the shape of distributions have to be taken into account properly. Reducible backgrounds as $\gamma j$ and $jj$ also demand this kind of careful treatment.

We show in the Figure~(\ref{dcos}) the $\cos(\tta)$ distribution for photons pairs decay of a spin-0 and a spin-2 particle at the 8 TeV LHC. Without detector effects, the spin-0 state decays isotropically in the CM system of the photon pair, resulting in a flat $\cos(\tta)$ distribution, while the spin-2 KK-graviton exhibits the quartic polynomial behavior of Equation~(\ref{P2grav}). After a full simulation, on the other hand, the distorted shapes of the distributions get more similar to each other. 
Given this more realistic situation, it is crucial to evaluate the real potential of the LHC to discriminate between the two cases by simulating the statistical fluctuations in a real experimental. This can be done performing a sufficiently large number of pseudo-experiments based on Monte Carlo (MC) simulations. 

We have imposed the following selection cuts based on the ATLAS analysis~\cite{atlas} in the generation of the signal and background MC events for the 7 and 8 TeV energy regimes at the LHC 
\begin{eqnarray}
& & E_{T_1} > 40\gev \;\;, \;\; E_{T_2} > 30\gev \nonumber \\
& & |\eta_{1,2}| < 2.37\;\;, \;\; \Delta R_{1,2} > 0.4 \nonumber\\ 
& &  100\gev < M_{1,2} < 160\gev 
\label{selection}
\end{eqnarray}
and excluding the calorimeter barrel/end-cap transition region $1.37 < |\eta_{1,2}| < 1.52$. The leading(sub-leading) identified photon is required to have $E_T>40(30)\gev$ inside the calorimeter barrel and within the 100-160 GeV invariant mass $M_{1,2}$ window. The same criteria was applied when jets or charged leptons are misidentified as photons. 

The SM Higgs boson production cross section from gluon fusion were normalized by the improved NNLO-NNLL QCD calculation~\cite{nnlo-nnll} at $m_H = 125\gev$ at the 7 and 8 TeV LHC. We have also included a 95\% identification efficiency per photon. The Higgs boson decay width was calculated at NLO QCD+EW with \texttt{HDECAY}~\cite{hdecay}. As a final result, around 25(20) Higgs boson events/$\ifb$ are expected for the chosen mass at the 8(7) TeV LHC, which are similar to the expected numbers from the experimental collaborations~\cite{atlas,cms}.

As we are considering the RS model just as a convenient example of an alternative spin-2 scenario against with we test the spin-0 hypothesis, the gravitons are assigned the same mass and width of the Higgs boson, and the cross sections for the production and decay of a resonant graviton is normalized by the SM Higgs rates. This is the most conservative scenario for model discrimination.

The MC background events were normalized by the experimental number of events quoted by ATLAS~\cite{atlas} for $5.8(4.9)\ifb$ at the 8(7) TeV LHC for the selection criteria of Equation~(\ref{selection}), respecting the measured composition of 74\%, 22\%, 3\% and 1\% for the $\gamma\gamma$, $\gamma j$, $jj$ and Drell-Yan contributions for a total of 35002(23619) background events at the 8(7) TeV LHC.

\section{The {\it sWeight} Technique}

For the selection cuts adopted by the experimental collaborations, after taking into account all the efficiency factors, around 3 signal $\gamma\gamma$ events from a Higgs boson decay are expected for each 1000 background events at the 7 and 8 TeV LHC~\cite{atlas,cms} in the mass range $100$--$160\gev$. Measuring the shape of any kinematic distribution thus requires an optimal selection technique.

Similar signal to background (S/B) ratios are encountered, for example, in B-physics experiments~\cite{exps} where an excellent signal {\it versus} background separation is achieved by means of the {\it sWeight} technique~\cite{splot}. The method was employed recently to discriminate between spins and parities of ``Higgs-alikes''  using four charged lepton events from $ZZ$ decays~\cite{lykken} also for very small signal to background ratios. We give here just a brief description of the method.

\subsection{Brief description of the formalism}

The {\it sWeight} formalism relies on an unbinned extended maximum Likelihood approach to a data sample in which signal and background events are mixed. If $N_s$ signal events and $N_b$ background events (which can be the outcome of the composition of a number of background components) are expected on average for a given integrated luminosity, the log likelihood for a given data sample of $N_e$ events is given by
\begin{equation}
{\cal L}=\sum_{k=1}^{N_e}\ln\left[N_sf_s(x_k)+N_bf_b(x_k)\right]-N_s-N_b
\end{equation}
where $f_s(x_k)$ and $f_b(x_k)$ are the probability density functions of a discriminating variable $x$ evaluated at the $k^{th}$ data event. These pdfs are obtained from MC simulation for signal and backgrounds and must be well understood in order the method to be reliable. 

Given the $\hat{N}_s$ and $\hat{N}_b$ that maximizes ${\cal L}$, we compute the {\it sWeights} for signal ($n=s$) and background ($n=b$) species for the $k^{th}$ data event as
\begin{equation}
_sW_n(x_k) = \frac{\mathbf{V}_{ns}f_s(x_k)+\mathbf{V}_{nb}f_b(x_k)}{\hat{N}_sf_s(x_k)+\hat{N}_bf_b(x_k)}
\end{equation}
The covariance matrix $\mathbf{V}$ is obtained by inverting the matrix $\mathbf{V}_{ni}^{-1}=-\partial^2{\cal L}/\partial N_n\partial N_i$ which can be computed by numerical means.

After the computation of the {\it sWeights}, we can populate histograms of any distribution of interest $y$, that is not correlated with the discriminating variable $x$, according to
\begin{equation}
\bar{y}_n = \sum_{|y_k-\bar{y}_n|<\delta y/2} {_s}W_n(x_k)
\end{equation}
where $\bar{y}_n$ is the central value of a given bin of width $\delta y$, and $y_k$ the value of the $y$ variable for the $k^{th}$ event. The histograms constructed from this prescription are called {\it sPlots} and none previous knowledge of the target distribution $y$ is assumed.

The {\it sPlots} are proved to bear important properties upon which stands the power of the method. First of all, they converge asymptotically to the true binned distributions and reproduce, on average, the finite statistics histograms. Summing up the events from signal and background {\it sPlots} recovers the number of events in the data sample in a composition given by the signal and background number of events provided by the maximum likelihood fit, i.e, $\hat{N}_s+\hat{N}_b=N_e$, while the statistical uncertainty on each $y$-bin can easily computed as
\begin{equation}
\sigma(\bar{y}_n) = \sqrt{\sum_{|y_k-\bar{y}_n|<\delta y/2} {_s}W^2_n(x_k)}
\label{ssd}
\end{equation}

Moreover, the {\it sPlots} can be merged on an event-by-event basis in a straightforward way allowing us to combine analyses from different experiments and different runs, for example. The normalizations and error bars of the merged {\it sPlots} are guaranteed to reproduce the correct ones asymptotically. 

A very important requirement for the {\it sWeight} technique to work is a reliable parametrization of the discriminating distribution. It means we must choose a well understood variable both for signal and background events. In our case, the natural best choice is the $\gamma\gamma$ invariant mass distribution $M_{1,2}$. The LHC collaborations usually get good parametrizations using a Crystal Ball function or combinations of gaussian distributions to fit a resonant $M_{1,2}$ structure~\cite{atlas,cms}. For the backgrounds, usual parametrizations are provided by polynomials and simple exponential functions~\cite{atlas,cms}. 

We found that our MC signal events are well described by 6 parameters from the combination of two gaussian distributions: the means, widths, and normalizations. As in the experimental situation, the total background is well parametrized, in the invariant mass region considered here, by a simple exponential distribution with two parameters, one for the shape and the other for the normalization of the distribution. We show in Figure~(\ref{fig:parameter}) the MC points for our simulated events and the fitting curves for the resonant scalar boson invariant mass and for the total background. Note that the maximum of the resonance occurs for a somewhat smaller value of the invariant mass compared to the scalar mass of $125\gev$. This is the effect of the photons energy loss inside the detector which can be compensated by the experimental collaborations using a proper calibration for the photon energy.

It is also important to check that the discriminating and the target variables are not correlated. We show in Figure~(\ref{fig:correl}) density plots in the $M_{\gamma\gamma}$ {\it vs} $\cos(\tta)$ plane for our simulated events and 
the joint probability function $P(M_{\gamma\gamma},\tta)=P(M_{\gamma\gamma})\times P(\tta)$ supposing independent variables. The independence can be further corroborated by computing the covariance between the variables, which, in this case, is $0.07$ for the spin-0 and $0.04$ for the spin-2 sample corresponding to the figure.
\begin{figure}
\resizebox{0.4\textwidth}{!}{%
\includegraphics[height=.20\textheight]{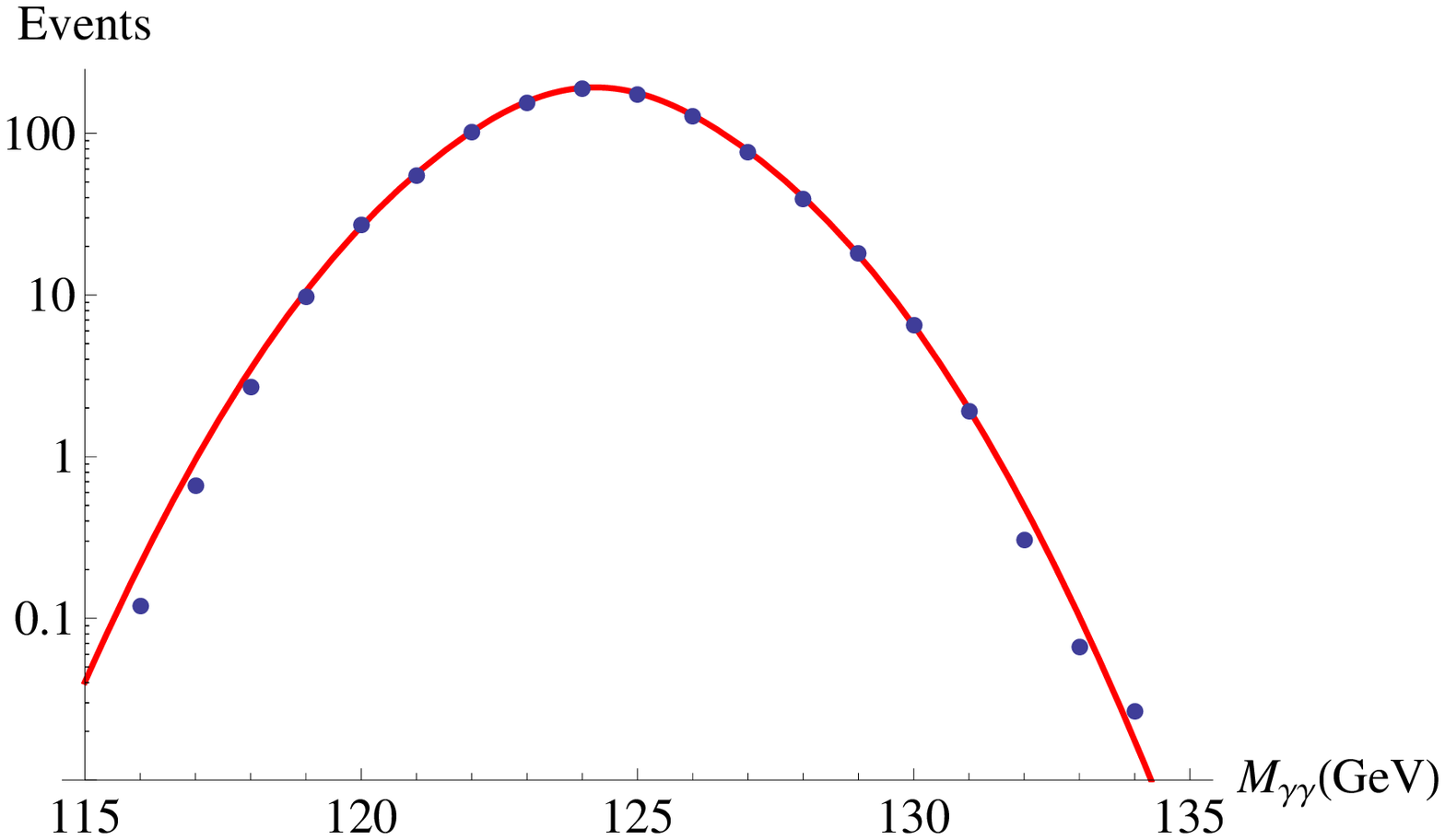}}
\resizebox{0.4\textwidth}{!}{%
\includegraphics[height=.20\textheight]{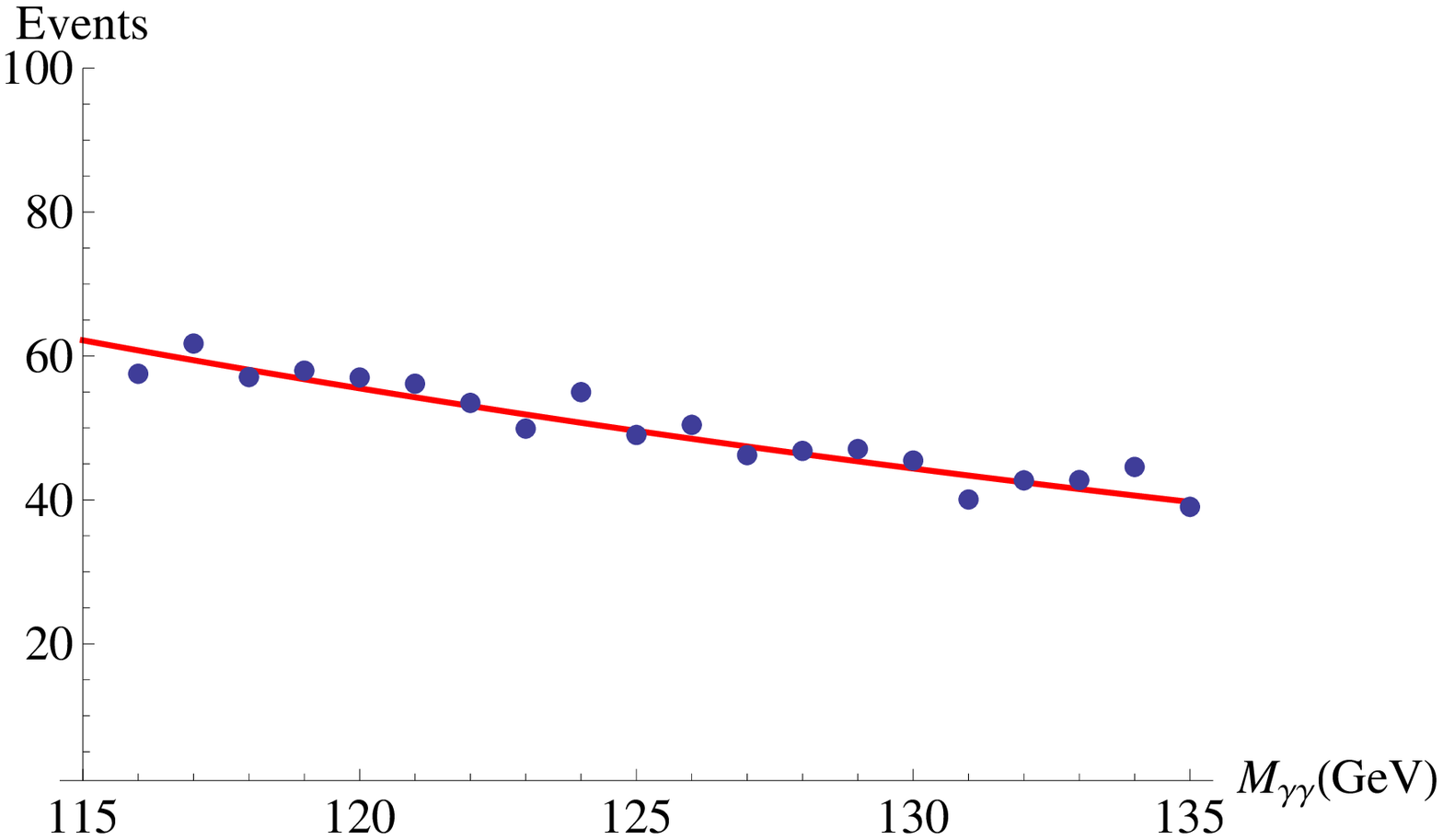}}
\caption{The points show the MC events for an $125\gev$ scalar resonance at the upper panel and for the total $\gamma\gamma$ backgrounds at the lower panel. The red curves are fitted from the MC points as a sum of two gaussian pdfs and an simple exponential pdf for the signal and the background $M_{\gamma\gamma}$ distributions, respectively.}
\label{fig:parameter}
\end{figure}

\subsection{{\it sPlots} of the scattering polar angle distributions}

From a large number of MC events generated, we randomly selected
$N_e$ pseudo-experiments in the statistical bootstrap sense, putting the events back into the sample in order they can be selected again. We verified that this procedure is equivalent to obtain random variates from the empirical distributions of the simulated samples. We show in the Figure~(\ref{fig:splots}) the $\cos(\tta)$ {\it sPlots} for the SM Higgs boson and the spin-2 KK-graviton. The figure also shows the histogram for a pseudo-experiment and the theoretical distribution as given by the MC simulation.

 It is truly amazing how well the {\it sPlots} reproduce the pseudo-experiments represented by the histograms in Figure~(\ref{fig:splots}) and the statistical errors associated to each bin. We superimpose the theory predictions including detector effects from a large number of MC events. The agreement between the theoretical curve and {\it sPlots} is very good. We also checked that the background expectations are very well described by the corresponding {\it sPlots}.

We present in the next section the statistical significances for spin-0 {\it vs} spin-2 model discrimination.
\begin{figure}
\resizebox{0.45\textwidth}{!}{%
\includegraphics[height=.20\textheight]{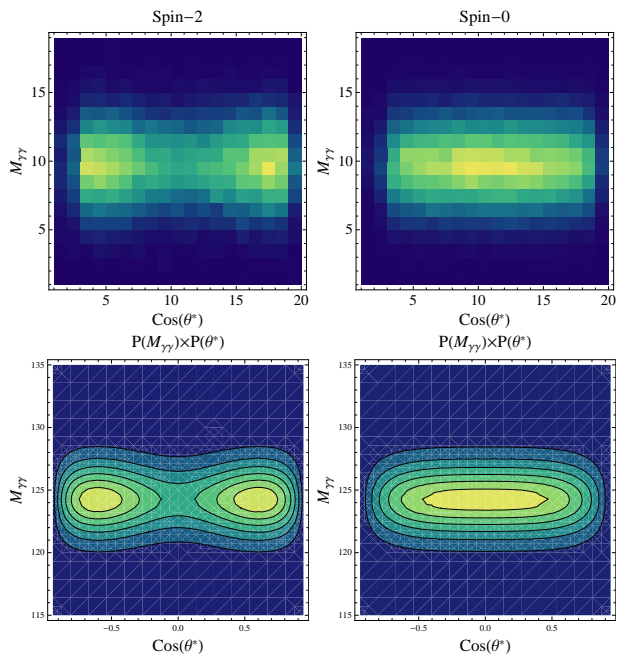}}
\caption{Upper plots: the empirical joint distribution $P\left(M_{\gamma\gamma},\tta\right)$ of generated MC events for a spin-0 (right panel) and a spin-2 resonance (left panel). Lower plots: the corresponding theoretical pdfs assuming that $M_{\gamma\gamma}$ and $\cos(\tta)$ are statistically independent variables.}
\label{fig:correl}
\end{figure}
\section{Statistical Significances for the Spin Discrimination}

 Given the optimal separation between signals and backgrounds, we are in the position to construct a statistic to perform the hypotheses tests in order to project the necessary amount of data to discriminate between a scalar boson and a KK-graviton alternative for a fixed confidence level (CL). 

The log likelihood ratio (LLR) is the best statistic to perform parametric tests and has been successfully used in recent works for the Higgs boson case~\cite{lykken,gritsan1,gritsan2} in $ZZ$ and $WW$ decays. Suppose that $\mu_0^{i}$ is the expected number of spin-0 events in a given bin $i$ of an {\it sPlot} histogram of $\cos(\tta)$ with $N_{bins}$ , and the corresponding number for a spin-2 distribution is $\mu_2^i$. If the data are denoted by $d^i$, the  log likelihood ratio to discriminate between the spin hypotheses is defined as
\begin{equation}
\Lambda = 2\sum_{i=1}^{N_{bins}}\left[\mu_0^i-\mu_2^i-d^i\ln\left(\frac{\mu^i_0}{\mu_2^i}\right)\right]
\end{equation}
\begin{figure}
\resizebox{0.45\textwidth}{!}{%
\includegraphics[height=.20\textheight]{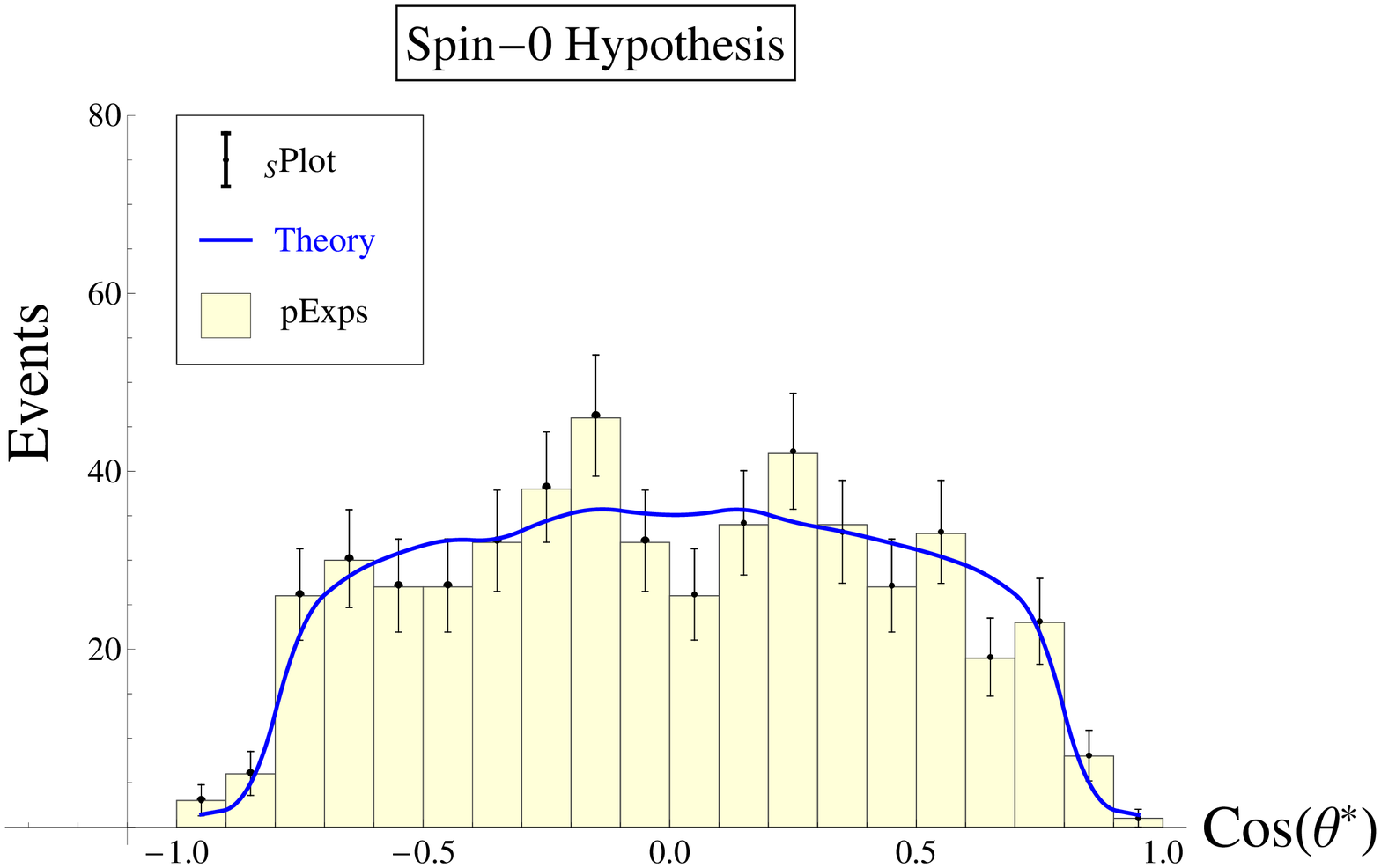}}
\resizebox{0.45\textwidth}{!}{%
\includegraphics[height=.20\textheight]{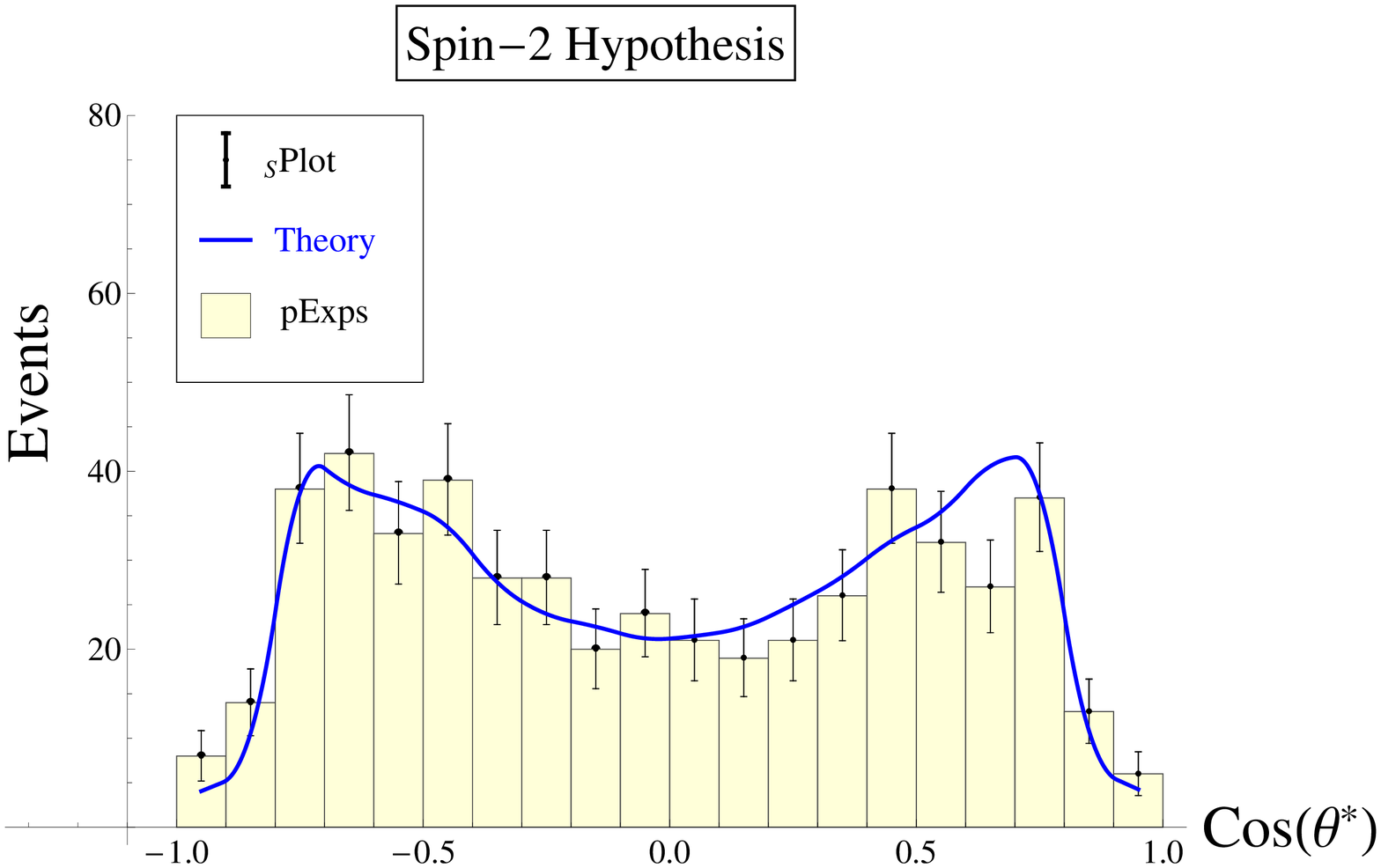}}
\caption{The $\cos(\tta)$ distribution for a scalar (upper panel) and a KK-graviton (lower  panel) including cuts, hadronization and detector effects. The histograms represent the outcome of a pseudo-experiment, the points with error bars are the {\it sPlots}, and the solid lines the theory prediction from a large number of generated MC events from \texttt{MadGraph5}.}
\label{fig:splots}
\end{figure}
With no free parameters, the LLR follows a gaussian distribution~\cite{cousins} whose mean scales with the integrated luminosity as $L$ and the standard deviation as $\sqrt{L}$.
For diphoton events, the scattering polar angle is the most useful discriminant between the alternative scenarios. Relying on just one distribution to construct the statistic, the LLR is shown to have similar discrimination power of a $\chi^2$ test~\cite{brandt}. The LLR performs as better as more discriminating variables are used together taking into account their mutual correlations~\cite{lykken}. Moreover, as we will show, it is rather sensitive to systematic errors. 

Another simple and robust statistic, although less powerful than the LLR, is the Center-Edge asymmetry defined as
\begin{equation}
A = \frac{\sigma(\lvert\cos(\tta)\rvert > 0.5)-\sigma(\lvert\cos(\tta)\rvert < 0.5)}{\sigma(\lvert\cos(\tta)\rvert > 0.5)+\sigma(\lvert\cos(\tta)\rvert < 0.5)}
\label{ce}
\end{equation}

From the theory point of view, an asymmetry is a model prediction which is intrinsically related to the shape of a given variable. In our case, the variable reflects the actual spin quantum number of the resonance. 

A sample of asymmetries follows a gaussian distribution with mean $\overline{A}$ and standard deviation $\overline{\sigma}_A$, the mean of the asymmetry error~\cite{lyons}
\begin{equation}
\sigma_A = \sqrt{\frac{1-A^2}{N}}
\label{sdce}
\end{equation}
where $N$ is the total number of events for a given integrated luminosity $L$.

Note that an asymmetry does not depend upon $L$, only its related statistical error. In fact, any systematic uncertainties affecting the signal yield as a multiplicative factor cancel in the ratio. There are several errors of this type, including systematics on luminosity, selection efficiencies, theoretical errors from renormalization/factorization scale uncertainties, for example. The most dangerous types of errors for an asymmetry measure are those related to bin migrations and errors that distort the shape of the distributions in general as the jet energy scale and parton distribution functions uncertainties, for example. Moreover, as it gathers all events in two larger sets -- the Center bin $\lvert\cos(\tta)\rvert < 0.5$ and the Edge bin  $\lvert\cos(\tta)\rvert > 0.5$ -- the central asymmetry value is less sensitive to statistical fluctuations.

On the other hand, if there is a contamination from background events, even when it is small, the asymmetries may be washed away or biased to a common value which requires a larger number of events for a good discrimination. Kinematic cuts are usually employed to get a cleaner sample of signal events, but this time a sculpting effect may distort the shape of the target variable and turns the predictions more similar. The {\it sWeight} technique circumvents these difficulties and allows us to fully explore the differences in the shapes of variables to obtain accurate asymmetry values.

From the point of view of the computational burden necessary to give a reliable estimate of the statistical confidence level of the model discrimination, a relatively small number of pseudo-experiments is necessary to build the probability distribution function of the asymmetries. We verified that $10^3$-$10^4$ events are sufficient to obtain the Gaussian parameters of the $A$ distribution with a very good precision.

Once the asymmetries for both alternatives have been calculated from their corresponding {\it sPlots}, and as we know that $P_0(data|A)$ and $P_2(data|A)$, the pdf of the spin-0 and spin-2 asymmetry~\footnote{Here, $data$ denotes the generated pseudo-experiments.}, respectively, are normally distributed, the statistical significance of the hypothesis test can be easily computed
\begin{equation}
Z(\sigma)=\Phi^{-1}(1-p)=\frac{\lvert \overline{A}_0-\overline{A}_2\rvert}{\overline{\sigma}_2}
\end{equation}

If $A_c$ is the chosen critical value for the test, the p-value $p$ is the probability of a type-I error ($\alpha$) given by $p=\int_{A_c}^\infty P_2(data|A)(x)dx$, and represented by the yellow shaded area ($p_1$) under $P_2(data|A)$ for $A_c=\overline{A}_0$ in Figure~(\ref{fig:pdfasym}). The inverse of the cumulative distribution function of the standard normal, $\Phi^{-1}$, calculated at $1-p$, gives the statistical confidence level $Z(\sigma)$ of the test in units of standard deviations of a standard gaussian distribution. 

If we have good reasons to believe that the $125\gev$ resonance is indeed a scalar particle, the critical value is set to the spin-0 mean or median prediction, $A_c=\overline{A}_0$ and $p=\alpha=p_1$ is the probability to accept the spin-0 hypothesis when, in fact, it is wrong. If we require the p-value to be very small, we loose control on the probability of rejecting the scalar hypothesis when it is correct one -- this is the type-II error ($\beta$), which can be calculated as $\beta=\int_{-\infty}^{\overline{A}_0} P_0(data|A)(x) dx=0.5$. This is a prejudicious choice based on {\it a priori} belief in the scalar option. 


A choice with no preferences would be taking a critical value for which the type-I and type-II errors were equal~\cite{cousins}. That is it, if no prior information would be available at the time of the spins discrimination, there would not be any reason to favor a particular choice. In this case, the p-value increases by the magenta shaded area ($p_2$) of the Figure~(\ref{fig:pdfasym}) while $\beta$ decreases to $p_3+p_4 < 0.5$. If the pdfs are not too asymmetric, a good approximation to the critical value is $A_c=(\overline{A}_0+\overline{A}_2)/2$ and the confidence level in any of the two hypotheses is smaller by a factor of 2.

Under the light of the current information about the new boson, we believe that the prejudicious choice of a spin-0 particle hypothesis is not conceptually misleading. Any serious alternative spin-2 candidate should show up with similar branching fractions to pairs of $Z$, $W$, $\gamma$, and SM fermions, beside the right production rate.

\subsection{Incorporating systematic errors}
\begin{figure}
\resizebox{0.4\textwidth}{!}{%
\includegraphics[height=.20\textheight]{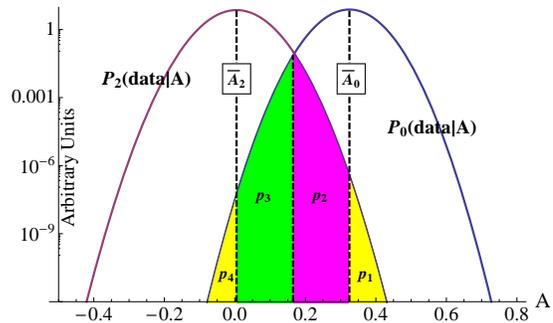}}
\caption{The probability density function of the center-edge asymmetry $A$ extracted from the fitting to a sample of $10^4$ pseudo-experiments for the spin-0 and spin-2 distributions. The colored areas represent the type-I and type-II errors for different critical values.}
\label{fig:pdfasym}
\end{figure}
Once we have reliably estimated the signal and background pdfs, it is possible to incorporate sources of systematic errors in the computation of the statistical confidence level in an approximate fashion. It is beyond the scope of this work to estimate all sources of systematic uncertainties that could has an effect on the calculation of asymmetries, which is a task aimed to the experimental collaborations. Yet, we provide a simplified analysis to show that we should not expect any large deviations due systematic uncertainties on the asymmetry measures, but the tests based on the LLR are somewhat degraded.

Suppose the pdfs depend on a number of nuisance parameters $\{ \varepsilon_1,\cdots,\varepsilon_n \}$ that describe uncertainties from $n$ sources of systematic errors. To embody these effects in the attainment of the confidence belt, we could marginalize the nuisance parameters in a Bayesian approach, convoluting the asymmetry pdf with a prior density $\pi(\varepsilon)$ for each systematic uncertainty. For example, for a single nuisance parameter normally distributed around $\varepsilon$ with error $\sigma_\varepsilon$~\cite{CH}, the asymmetry probability density function is given by
\begin{equation}
P(data|A,\varepsilon)=\int_{0}^{+\infty} P(data|A,\varepsilon^\prime)\frac{e^{\frac{(\varepsilon^\prime-\varepsilon)^2}{2\sigma_\varepsilon^2}}}{\sigma_\varepsilon\sqrt{2\pi}}d\varepsilon^\prime
\label{sys}
\end{equation}
In fact, a hybrid technique can be used~\cite{pekka}. A frequentist approach is possible in the sense that the effects of these systematic errors could be estimated from an ensemble of pseudo-experiments each of them with a randomly selected value of $\varepsilon$. 

As we are interested in signal events $\mu_s$ as simulated from theory, unlike the backgrounds, we integrate over nuisance parameters of the signal pdfs only.  For the correlated errors that we choose to illustrate the impact of systematic errors -- the uncertainty on the integrated luminosity ($\varepsilon_L$) and the selection efficiency ($\varepsilon_{eff}$) -- it is easier to just adopt the frequentist approach and multiply the number of signal events as $\mu_s^\prime = \mu_s(1+\varepsilon_L)(1+\varepsilon_{eff})$ for each generated pseudo-experiment. We adopt the uncertainties quoted in the ATLAS experiment~\cite{atlas}: a relative uncertainty of 3.6\% in the luminosity and 11\% in the selection efficiency.

 In fact, none of these correlated sources of systematic errors changed the value of the original asymmetries noticeably nor their probability density functions, confirming the expectation that this kind of uncertainty cancel in the ratio. On the other hand, the probability density functions of the LLR statistic broadens, increasing the p-values. We compare the performance of the center-edge asymmetry and the LLR statistic in the next section.


\section{Results and Discussion}

We present in Table~I the values of the asymmetries for the spin-0 and the spin-2 resonances computed from the parton level MC events, the distorted detector level MC events, and from the corresponding {\it sPlots} with marginalized correlated systematic errors as discussed in the previous section.

An approximate shift of $0.3$ in the asymmetry values is observed from parton level to detector level and {\it sPlots} both for spin-0 and spin-2 scenarios. This is the consequence of the migration of events from the edges to the center of the $\cos(\tta)$ distributions due the geometry restriction of the detectors.

As we pointed out earlier, the {\it sPlots} from any number of channels can be added straightforwardly. We can take advantage of this and combine all the available data produced in ATLAS and CMS collaborations and from the 7 and 8 TeV runs. 

Figure~(\ref{stat}) shows the main results of this work. The statistical significance of the spin-0 boson {\it versus} spin-2 KK-graviton discrimination, based on the asymmetries of their $\cos(\tta)$ distributions, are shown for the 8 TeV ATLAS run solely, the 7 and 8 TeV runs, and for the entire data set from ATLAS and CMS for both runs. 
\begin{table}
\begin{center}
\begin{tabular}{c|c|c|c}
\hline
\hline
 & Parton level & Detector level & {\it sPlots} \\
\hline
spin-$0^+$ & $0$ & $0.324$ & $0.325\pm 0.042$\\
\hline
spin-$2^+$ & $-0.207$ & $0.005$ & $0.005\pm 0.044$ \\
\hline
\hline
\end{tabular}
\end{center}
\caption{The values of the center-edge asymmetries for a spin-0 and a spin-2 resonance at parton level, detector level, and from {\it sPlots} with $10^3$ pseudo-experiments.}
\label{tab1}
\end{table}

The upper panel of Figure~(\ref{stat}) show the statistical significance in the spin-0 hypothesis against the spin-2 hypothesis. The solid line represents $Z(\sigma)$ as a function of the integrated luminosity at the 8 TeV LHC for the ATLAS experiment only. A $5\sigma$ discrimination is possible with $\sim 15\ifb$. Adding the 7 TeV run increases the significance to $5.5\sigma$ with the same amount of data from the 8 TeV run.

For this luminosity, however, a $5\sigma$ signal to background discrimination cannot be reached relying only on the $\gamma\gamma$ channel~\cite{atlas,cms} in the case of a SM Higgs boson. It should be pointed out, however, that the present observed significance is $4.5\sigma$ which characterizes an $1.8$ signal strength in the $H\to\gamma\gamma$ channel. If this enhancement persists, with the current signal strength, a $5\sigma$ spin discrimination of the spin-0 hypothesis against the spin-2 with minimal couplings should already be feasible.

\begin{figure}
\resizebox{0.4\textwidth}{!}{%
\includegraphics[height=.20\textheight]{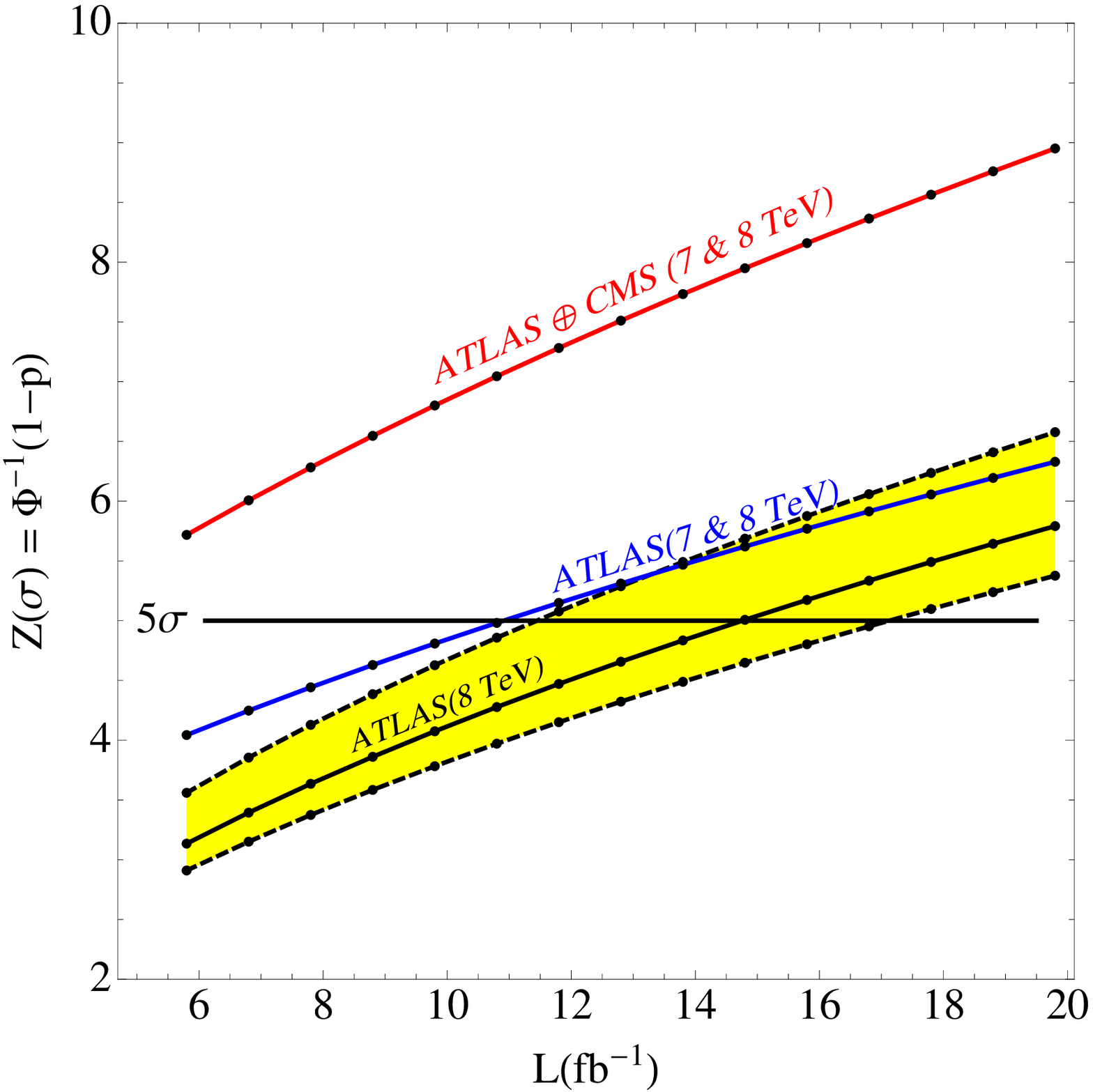}}
\resizebox{0.395\textwidth}{!}{%
\includegraphics[height=.18\textheight]{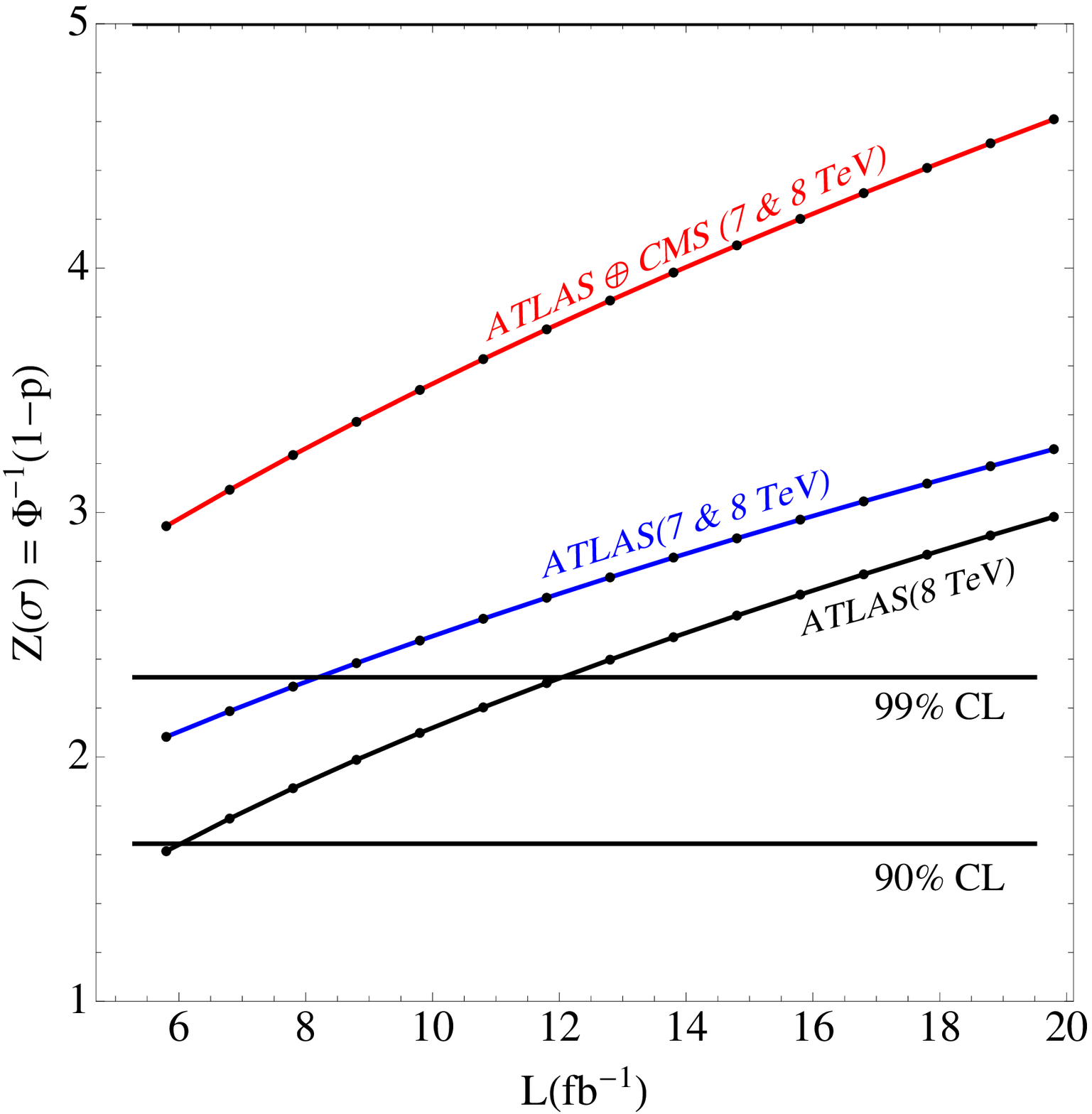}}
\caption{The expected significance for spin-0 {\it versus} spin-2 KK-graviton hypothesis, based on the asymmetry measure, at the 8 TeV LHC for the ATLAS experiment (lower solid lines), 7 and 8 TeV combined data for ATLAS (middle solid lines), and combining all available data from the 7 and 8 TeV runs for ATLAS and CMS (upper solid lines). The dashed lines at the upper panel represent the discrimination power of a LLR statistic with systematic uncertainties ranging from 0 (upper line) to 30\% (lower line) in the normalization of the signal strength. The lower panel depicts the significance calculated in the non prejudicious approach where $\alpha=\beta$ as explained in the text. The 90\% and 99\% CL lines are also shown.}
\label{stat}
\end{figure}
If we conservatively assume that the current excess is just a momentary statistical fluctuation, the $5\sigma$ C.L. in the signal {\it vs} background hypothesis will be reached for around $40\ifb$
of data, or some $20\ifb$ per experiment after ATLAS and CMS combination for a SM signal strength~\cite{atlas,cms}, but we still can combine the ATLAS and CMS data for the 7 and 8 TeV runs for the spin diagnosis. In this case, the red line in the upper panel of Figure~(\ref{stat}) represents the combined result with a luminosity $L$ for each experiment. For $L=20\ifb$ a very high statistical C.L. of almost $9\sigma$ in the scalar hypothesis can be reached. Note that with the current data, a combined analysis is already able to discriminate between both spin alternatives. Of course, as the signal excess in the diphoton channel can not be attributed to a new resonance at the required statistical level, for the LHC standards, in principle, it makes no sense to talk about which kind of particle is being produced.

In spite of being usually a more powerful statistic, the log likelihood ratio is much more sensitive to systematic uncertainties. To evaluate the impact of these uncertainties, we adopted a larger variation in a single correlated systematic error in the signal prediction, varying its magnitude from 0 to 30\%, and distributed normally as discussed in the previous section. 
The shaded yellow band, between the dashed lines in the upper panel of Figure~(\ref{stat}), shows the loss in the discrimination power in the LLR statistic. 

With no systematic errors, the LLR performs around 10\% better than the asymmetry for ATLAS 8 TeV run, which corresponds to the upper dashed line. As the uncertainty increases, the LLR pdfs broaden and there is a loss in the discerning power of the test. Eventually, the LLR statistic is expected to perform worse than the asymmetry measure, the solid line inside the shaded region. The lower dashed line depicts the significance achieved using the LLR and a larger systematic uncertainty of 30\% in the normalization of the signal yield. Similar behavior was found for more data from combinations of runs and experiments.

By its turn, the variation in the discrimination power of the asymmetries is less than 1\% for the same 0-30\% variation in the magnitude of the single systematic uncertainty.

The lower panel of Figure~(\ref{stat}) shows the significances by choosing a critical value for which the type-I and type-II errors are nearly equal in the hypothesis test. This is the no prejudice scenario discussed in section IV. Here, 4 times more luminosity is necessary to reach the same C.L. compared to the prejudicious scenario. Combining all the available data from the two experiments will enable a $5\sigma$ discrimination with $25\ifb$ per experiment and around $4.5\sigma$ for $20\ifb$. On the other hand, a 99\% CL discrimination is well within the LHC possibilities for the 8 TeV run.  

\section{Conclusions}
 
The picture delineated by the current data on the new $125\gev$ boson discovered at the LHC already points to a Higgs boson. Yet, its identity, which must be uniquely assigned by the set of its quantum numbers, has not been determined -- this identification is a crucial step to confirm the new particle as a Higgs boson, be it SM or not. Discriminating its spin number will break the boson degeneracy into a scalar or a tensor state, once the vector alternative has already been discarded from the observed decays into photons pairs.

The $\gamma\gamma$ channel is even more useful in the spin determination. We showed that by measuring a center-edge asymmetry in the {\it sPlots} of the  scattering polar angle of the diphotons, is sufficient to discriminate between a scalar $J^P=0^+$ and a tensor $J^P=2^+$ resonance minimally coupled to photons and gluons, using the KK-gravitons of the RS model as a straw man scenario. By the way, although there is a large number of possibilities of non minimal couplings, the KK-graviton inspired couplings are not the easiest conceivable scenario against whom the scalar hypothesis could be tested, from the point of view of the amount of data required to a high discrimination level.

The asymmetries were also shown to be rather insensitive to systematic uncertainties, contrary to the usually more powerful log likelihood ratio statistic, and performs even better than the LLR for errors of moderate magnitude. 

The {\it sWeight} technique proved to be essential in this good performance, once it allows one to optimally separate signal and background events, which ultimately lead to a maximum discerning power based on the shapes of the distributions, being only limited by detector effects. The {\it sPlots} can be combined in a straightforward way and all the information accumulated in the 7 and 8 TeV runs from both ATLAS and CMS can be readily used in the analysis.

As a result of the application of this technique, we found that by the time the $\gamma\gamma$ channel alone favors the signal hypothesis, with a SM strength, against the background one at a $5\sigma$ confidence level, the scalar {\it versus} graviton hypotheses will be discriminated to a higher significance level of $\sim 9\sigma$. A $5\sigma$ spin discrimination will possible, though, with less data, and an integrated luminosity of approximately $10\ifb$ for ATLAS or CMS is sufficient to discriminate between the alternatives. Also, if a large signal strength in the $\gamma\gamma$ channel, comparable to the current one, is really confirmed, the spin-0 hypothesis is already favored to $\sim 5\sigma$ combining the ATLAS and CMS data.

Keeping both the probability to accept the wrong hypothesis and the probability to reject the right one under control, in a less prejudicious approach, requires more data for a $5\sigma$ C.L. discrimination. We estimate that $25\ifb$ for each experiment would be necessary in this case. Relying on $20\ifb$ per experiment a $\sim 4\sigma$ significance is still possible.

\section{Appendix\label{sec:ApA}} 

We present here the general amplitude for the decay of a spin-2 particle into a pair of photons as given by Equation~(\ref{coup}). The case for a decay into a massive pair of gauge bosons can be found in Refs.~\cite{gritsan1,gritsan2}.

The operators ${\cal O}_i\; ,\; i=1,\cdots, 5$ are written in terms of the spin-2 wave function -- a traceless symmetric tensor $t_{\mu\nu}$, the photons momenta $q_{1\mu}$ and $q_{2\nu}$, and the polarization vectors of the photons $\epsilon^\mu_{1,2}$. We show the explicit form of these operators just below
\begin{eqnarray}
{\cal O}_1 & = & 2g_1t_{\mu\nu}f_1^{*\mu\alpha}f_2^{*\nu\alpha} \\
{\cal O}_2 & = & 2g_2t_{\mu\nu}f_1^{*\mu\alpha}f_2^{*\nu\beta}Q_\alpha Q_\beta \\
{\cal O}_3 & = & g_3t_{\beta\nu}(f_1^{*\mu\nu}f_{2\mu\alpha}^*+f_2^{*\mu\nu}f_{1\mu\alpha}^*)\tilde{Q}^\alpha \tilde{Q}^\beta \\
{\cal O}_4 & = & g_4f_1^{*\alpha\beta}f_{2\alpha\beta}^* t_{\mu\nu}\tilde{Q}^\mu \tilde{Q}^\nu \\
{\cal O}_5 & = & g_5f_1^{*\alpha\beta}\tilde{f}_{2\alpha\beta}^* t_{\mu\nu}\tilde{Q}^\mu \tilde{Q}^\nu
\end{eqnarray}

The vectors $Q_\mu$ and $\tilde{Q}_\nu$, and the tensors $f^{\mu\nu}$ and $\tilde{f}^{\mu\nu}$ are also combinations of the photons momenta and polarization vectors 
\begin{eqnarray}
Q_\mu = q_{1\mu}+q_{2\mu} \;\;\; &,&\;\;\; \tilde{Q}_\mu = q_{1\mu}-q_{2\mu} \\
f_i^{\mu\nu} = \epsilon_i^\mu q_i^\nu-\epsilon_i^\nu q_i^\mu \;\;\; &,&\;\;\; \tilde{f}_{i\mu\nu} = \epsilon_{\mu\nu\alpha\beta}\epsilon_i^\alpha q_i^\beta
\end{eqnarray} 

The traceless vector $t_{\mu\nu}$ is transverse to its momentum $t_{\mu\nu}Q^\mu=0$ just like the polarization vectors and the momenta of the photons $\epsilon_{i\mu}q_i^\mu=0$.

The coupling $g_5$, and the corresponding operator, corresponds to the coupling $g_8$ of the scattering amplitude from the Equation (18) of Ref.~\cite{gritsan2}. Note that this is a parity-odd contribution that represents the pseudo-tensor $J^P=2^-$ case.

\acknowledgments
I would like to thank Oscar Eboli for his careful reading of the manuscript and valuable suggestions and the Departamento de F\'\i sica Matem\'atica da Universidade de S\~ao Paulo where part of the simulations of this work were done.

\end{document}